\documentclass[aps,prb,twocolumn,superscriptaddress]{revtex4-1}

\usepackage{longtable}
\usepackage{amsfonts}
\usepackage{graphicx}

\begin{document}
 \title{Local breakdown of the quantum Hall effect in narrow single layer graphene Hall devices}

\author{Cenk Yan{\i}k}
\author{Ismet I. Kaya}
 \email{iikaya@sabanciuniv.edu}
 \affiliation{Faculty of Engineering and Natural Sciences, Sabanci University, Tuzla, 34956 Istanbul, Turkey}
 \affiliation{Sabanci University Nanotechnology Research and Application Center, Tuzla, 34956 Istanbul, Turkey}
 \begin{abstract}
  We have analyzed the breakdown of the quantum Hall effect in 1~$\mu m$ wide Hall devices fabricated from an exfoliated monolayer graphene transferred on $SiO_x$.  We have observed that the deviation of the Hall resistance from its quantized value is weakly dependent on the longitudinal resistivity up to current density of 5 A/m, where the Hall resistance remains quantized even when the longitudinal resistance increases monotonously with the current. Then a collapse in the quantized resistance occurs while longitudinal resistance keeps its gradual increase. The exponential increase of the conductivity with respect to the current suggests impurity mediated inter-Landau level scattering as the mechanism of the breakdown. The results are interpreted as the strong variation of the breakdown behavior throughout the sample due to the randomly distributed scattering centers that mediates the breakdown.
\end{abstract}

\date{\today}

\maketitle

\section{Introduction}

Shortly after the discovery of the quantum Hall effect,\cite{vK} the physical limits of this effect in various conditions such as sample mobility, temperature and bias current were widely investigated in two dimensional electron gas (2DEG) structures.\cite{Nachtwei}
The breakdown of the quantum Hall effect, which is observed as the abrupt increase in the longitudinal resistance with an associated loss in the quantization of Hall voltage is the major obstacle against improving the precision of the electrical resistance standard.
Although a comprehensive model that explains all the experimental results  on the breakdown of quantum Hall effect does not exist, the proposed models provide insight to different aspects of the breakdown. Some of these models are  intra\cite{Ebert} and inter\cite{Eaves} Landau-Level transitions, electron heating,\cite{Komiyama86,Takamasu} impurity mediated tunneling and relaxation of the electrons\cite{Kaya98,Kaya99,Guven} and electron phonon interactions.\cite{Streda,Balev}

Graphene is a two dimensional conductor which has linear energy-momentum dispersion and much larger Landau-level (LL) separation than GaAs based 2DEG for the same magnetic field.\cite{Novos05}
This promises obtaining more precise quantization of the Hall resistance, or under less strict conditions such as higher temperatures\cite{Novos07} or lower magnetic fields.
Therefore, graphene is a good candidate for the quantum Hall resistance metrology towards better precision and wider applications.
A relative uncertainty of a few parts per billion (bpp) in the quantized resistance, $R_K=h/e^2$ can be achieved in GaAs based Hall devices.\cite{Jeckelmann,Piquemal}
Early attempts in exfoliated graphene yielded a relative uncertainty of 15 parts per million (ppm) in the quantized resistance.\cite{Giesbers}
Since then the measured uncertainty of $R_K$ in the graphene based samples has been improved.
Exfoliated monolayer and bilayer graphene samples yielded 0.5 ppm on $SiO_x$ surface\cite{Guignard} and $6$ pbb on GaAs surface.\cite{Wosz}
Recently 0.3 ppb accuracy has been achieved in a large area epitaxial graphene Hall device.\cite{Tzal,JanssenNJP}
Direct comparison of the resistance quantum in epitaxial graphene and in GaAs 2DEG has shown no difference within the relative standard uncertainty of 8.6 parts in $10^{11}$.\cite{JanssenPRB}
There is sufficient conviction that graphene can improve the precision in  $R_K$ and for this the breakdown of the quantum Hall effect in graphene needs to be better understood. The breakdown of the graphene has recently been experimentally studied in a few publications. Singh \textit{et al.} found the critical current in monolayer graphene around 1~A/m and interpreted the results as the presence of disorder-induced broadening of Landau levels and inhomogeneous charge distribution.\cite{Singh} On the other hand, significantly higher breakdown current density of 8~A/m has also been observed in exfoliated monolayer graphene and it has been related to the high energy loss rates of hot carriers in graphene. It was also predicted that in monolayer graphene the breakdown currents could be as high as 43~A/m.\cite{Baker}

In this work we investigated the breakdown phenomenon in narrow monolayer graphene Hall bar samples addressing the role of the charge inhomogeneity. Microscopic inhomogeneity of the dopant density in 2DEG causes potential fluctuations in the sample interior. These microscopic inhomogeneities lead to slight spatial variations in the carrier density and the current distribution in large area samples. Altough they are small, these fluctuations are enhanced near the breakdown of the quantum Hall effect in 2DEG and result in large spatial variation of the critical currents.\cite{Cage83} The charge impurities has been shown to significantly enhance the inter-LL scattering rates and reduce the critical current in monolayer and bilayer graphene samples.\cite{Guignard}

\section{Experimental Results and Discussion}

\begin{figure}[ht]
\begin{center}
  \includegraphics[width=0.45\textwidth]{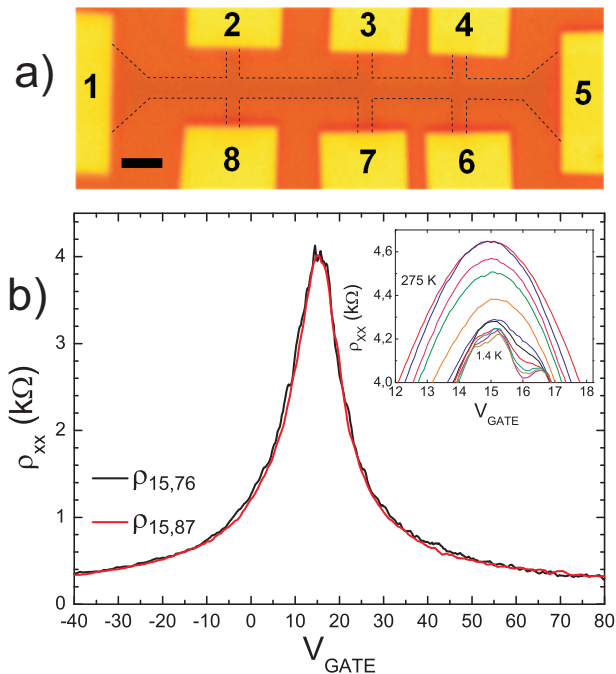}\
  \caption{\label{Fig1}
  a) The optical microscope image of the measured sample. Yellow pads are Cr/Au contacts defined by electron beam lithography. Current leads are marked as 1 and 5. During the measurements the electrons are injected from the lead 5. Dashed lines mark the borders of the graphene. Scale bar is 2~$\mu m$.
  b) The longitudinal resistivity, $\rho_{xx}$ versus the back gate voltage, $V_{GATE}$ measured between the contacts 6-7 ($L = 5~\mu m$) and 7-8 ($L = 7.5~\mu m$) ($\rho_{15,76}$, $\rho_{15,87}$) at $1.4$~K.
  Inset shows  $\rho_{15,76}$ versus $V_{GATE}$ for temperatures 275, 175, 135, 115, 77, 47, 37, 27, 10, 5 and 1.4 K.}

\end{center}
\end{figure}

The graphene flake used in the experiments is exfoliated from a highly ordered pyrolytic graphite (HOPG), transferred on to a silicon substrate coated with $285 nm$ of $SiO_x$ and is verified to be monolayer by Raman spectroscopy. Electron beam lithography and oxygen plasma etching are used to define the Hall  bar geometry with $w=$~1~$\mu m$ width as shown in Fig.~\ref{Fig1}(a). The current leads were given a funnel shape to gradually reduce the effect of the hot spots at the injection corners. The graphene arms were extended outside the Hall bar to form the potential probes to prevent the electrode-induced doping of the graphene. This geometry also minimizes the interference of the metal leads with the current flow and provides a well defined bar width through the sample.

 The contacts on the graphene are defined by electron beam lithography and Cr/Au (5/30~nm) evaporation. The sample is heat treated in $10^{-5}$~mbar vacuum by a combination of external heating up to $120\,^{\circ}{\rm C}$ and current heating by passing up to 1~mA dc current before the measurements. The heat treatment cycles were carried out inside a built-in chamber on top of the cryostat allowing the sample transfer without breaking of the vacuum. Then the DC measurements are done within a temperature range of 1.4-300~K. The degenerately doped silicon substrate served as the back-gate to tune the density of carriers.  After the heat treatment, the Dirac point was settled to +15~V which indicated that the graphene is p-type doped.  The source of the persistent doping even after such an aggressive heat treatment is probably due to residues trapped between the graphene and the substrate or the impurities from the $SiO_x$ substrate itself. We experienced in several samples that the current annealing was quite delicate and may result in highly inhomogeneous graphene if done excessively.  The Hall mobility of the device is measured to be 8300~$cm^2/V-s$ at $n=5\times10^{11}~cm^{-2}$ corresponding to a mean free path of $\textit{l} = 0.07~\mu m$. Since the mean free path and the size of the sample are comparable, there is an enhancement of the inhomogeneity over the Hall bar area. Fig.~\ref{Fig1}(b) shows the resistivity of the sample $\rho_{xx}=R_{xx}w/L$ as a function of the gate voltage measured between the two adjacent voltage probes (6-7 and 7-8) when current is passed between the leads 1-5. at $T = 1.4$~K. Both resistivity traces well coincide apart from the small fluctuations. These conductance fluctuations are repeatable and represents the varying effects of the random distribution of the scatterers on different potential probes. Inset in the Fig.~\ref{Fig1}(b) shows the emergence of the conductance fluctuations as the temperature is lowered below 77~K.

\begin{figure}[h]
\begin{center}
  \includegraphics[width=0.45\textwidth]{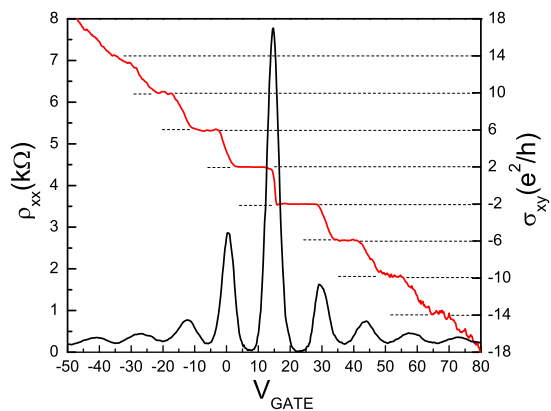}\
  \caption{\label{Fig2}
  Longitudinal resistivity $\rho_{15,76}$ (black curve) and the Hall conductance $\sigma_{15,46}$ (red curve) as a function of the gate voltage at $B=11~T$ and $T=1.4~K$ with a bias current $I = 0.7~\mu A$. Hall plateaus at the fillings factors $\nu=\pm2, \pm6, \pm10$ indicate that the sample is monolayer graphene.}
\end{center}
\end{figure}

In Fig.~\ref{Fig2} the Hall conductance, $\sigma_{xy}$ and the longitudinal resistivity, $\rho_{xx}$  are plotted as a function of the gate voltage at $B=11$~T and $T=1.4$~K.
The quantum Hall plateaus in $\sigma_{xy}$  corresponding to filling factors of $\nu=\pm2, \pm6$ and $\pm10$ are well defined and confirms that the sample is made of monolayer graphene.
$\nu=\pm2$ plateaus are the sharpest and the breakdown of the quantum Hall effect for the sample was studied in detail around these filling factors.

\begin{figure}[h]
\begin{center}
  \includegraphics[width=0.45\textwidth]{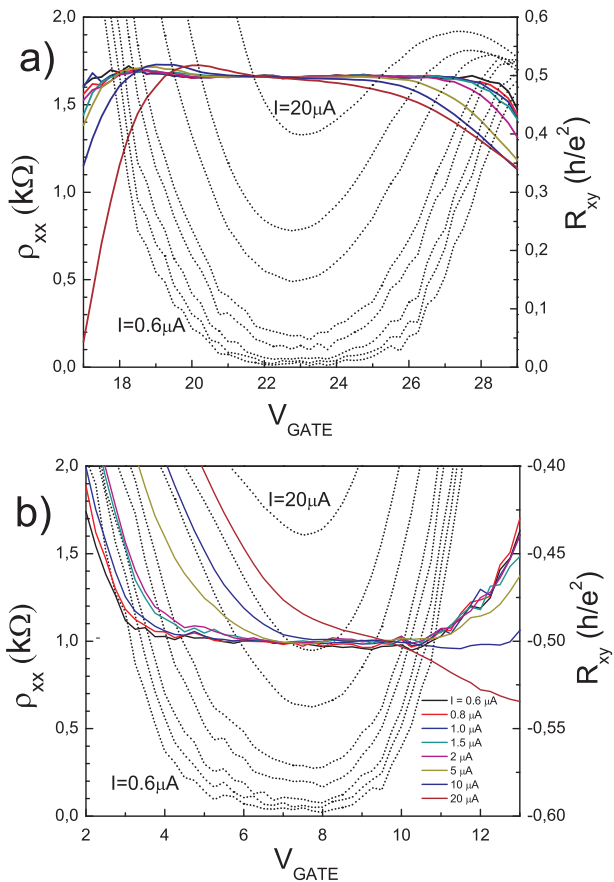}
  \caption{\label{Fig3}
  The evolution of the longitudinal resistivity $\rho_{15,76}$ (black dots) and the Hall resistance $R_{15,46}$ (colored solid lines) around the filling factors a) $\nu=2$ and b) $\nu=-2$ as a function of the gate voltage, $V_{GATE}$ at $B=11~T$, $T=1.4~K$ with currents I = 0.6, 0.8, 1, 1.5, 2, 5, 10, 20 ($\mu A$) as labeled in (b).}
\end{center}
\end{figure}

The breakdown of the quantum Hall effect is characterized simultaneously for the longitudinal resistance between the probes 6-7, 7-8 and the Hall resistance between the probes 4-6.
The rest of the voltage probes, had too high contact resistances to perform accurate measurements.
Fig.~\ref{Fig3}(a) displays the evolution of the longitudinal resistance minima, $\rho_{xx}$ and the Hall resistance plateaus, $R_{xy}$ around the filling factor $\nu= +2$ when the current is increased from 0.6 $\mu A$ to 20 $\mu A$ corresponding to maximum current density $j_x=I/w=20~A/m$ for the $w=1~\mu m$ sample width.
The results of the same kind of measurements for $\nu= -2$ are shown in Fig.~\ref{Fig3}(b).
In these graphs, the Hall resistance and the longitudinal resistance are measured between the contacts 4-6 and 6-7 respectively with a shared contact.
Therefore, $R_{xy}$ and $\rho_{xx}$ measurements probe the adjacent zones of the Hall bar; nevertheless not exactly the same as it is always the case for any measurement with the Hall bar geometry.
We observe quite different breakdown behavior for the Hall and longitudinal resistances for this specific sample and attribute it to local variations throughout the sample due to microscopic inhomogeneity of the unintentionally doped graphene.

We observe that both $\rho_{xx}$ and $R_{xy}$ follow a normal breakdown behavior.
As the current increases, longitudinal resistance minima increase from their nearly nondissipative value and reaches $\rho_{xx} > 1~k\Omega$ at $I = 20~\mu A$ while the flat plateaus of the $R_{xy}$ gradually shrink and eventually vanish.
The conductance fluctuations are still visible in these $\rho_{xx}$ and $R_{xy}$ plots for relatively low currents, but with the increasing current they diminish and disappear rather erratically.
Some of the fluctuations shift and may even enhance with the increased current. This behavior is attributed to the shifting of the electron-hole puddles in the graphene with the expansion of local dissipative regions.\cite{Martin}
A close examination of the graphs indicates that the emergence of the breakdown clearly occurs at quite different current values for $\rho_{xx}$ and $R_{xy}$.
While $R_{xy}$ maintains its flatness up to 5-10 $\mu A$ range, $\rho_{xx}$ nearly reaches 1~$k\Omega$ for such high currents and already exceeds 100~$\Omega$ around 1~$\mu A$.

\begin{figure}[h]
\begin{center}
  \includegraphics[width=0.45\textwidth]{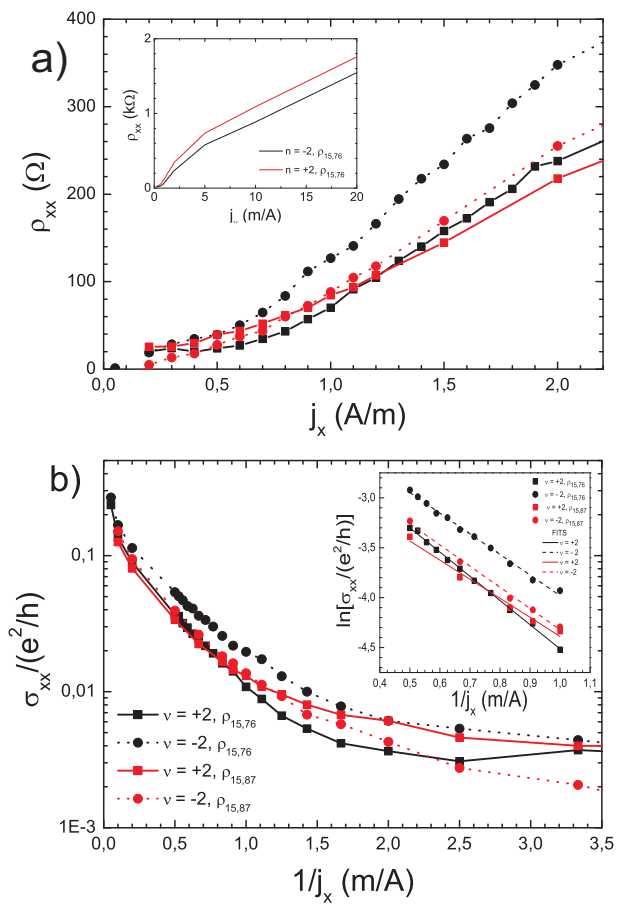}
  \caption{\label{Fig4}
  a) $\rho_{xx}$ versus current density, $j_x$ for $\nu = \pm 2$ and between the contacts 6-7 and 7-8. Longitudinal resistivity makes a rather gradual transition into dissipative regime with the increased current.
  Critical currents at filling factors $\nu=\pm2$ between the probes (6-7),(7-8) at $B=11~T$, $T=1.4~K$. Inset shows $\rho_{15,76}$ vs $j_x$ for $\nu = -2$ with the full range of currents.
  b) $\sigma_{xx}$ versus $1/j_x$ in semilog scale. Inset shows fittings at filling factors $\nu=\pm2$ between the probes (6-7),(7-8) at $B=11~T$, $T=1.4~K$ for the region $j_x=1-2~A/m$.}
\end{center}
\end{figure}

Fig.~4(a) displays the longitudinal resistivity between the voltage probes 6-7 and 7-8 as a function of the current density, $j_x$ for the low current range and the full range of up to 20~A/m.
Unlike the familiar breakdown behavior in 2DEG samples, the longitudinal resistivity makes a rather gradual transition into the dissipative regime with the increased current. This behavior is similar to those can be seen in the previous publications with exfoliated graphene.\cite{Guignard,Singh,Baker}
Although there is a nonlinear increase of resistivity below 1~A/m, an abrupt jump is missing in all of the traces.
Therefore, it is quite difficult to define the critical current for the breakdown.
Here we set a threshold resistivity of 40~$\Omega$ to determine the critical current values, $I_c$ for $\rho_{15,76}$ as 0.75~A/m ($\nu =+2$) and 0.50~A/m ($\nu =-2$); for $\rho_{15,87}$ as 0.50~A/m ($\nu =+2$) and 0.65~A/m ($\nu =-2$). A relatively high resistivity is chosen as the threshold resistance to reduce the error from the conductance fluctuations in this rather narrow sample.
We attribute the large variation in the critical current between the probes as well as between the filling factors to the inhomogeneity of the sample.
High local electric fields due to the applied gate voltages in the sample should effect the charge distribution.
For the filling factors $\nu=+2$ and $\nu=-2$ the gate voltage has the values $V_{GATE} = 23~V$ and $V_{GATE} = 7.5~V$.
Even for relatively clean samples due to the electron-hole puddles in graphene, gate voltage may affect the charge distribution through the sample and it can cause discrepancy for the values of critical currents.

Fig.~4(b) displays the semilog plot of the $\sigma_{xx}$ versus $1/j_x$.
AAn exponential variation of $\sigma_{xx}$ between $j_x=0.6-5~A/m$ can be seen in this graph for all the traces although their current range and slope slightly vary.
For lower currents longitudinal conductivity saturates to its minimum value originating from the conductance fluctuations.
For the high current densities ($j_x > 5~A/m$) $\sigma_{xx}$ increases faster with the current.
Inset in Fig.~\ref{Fig4}(b) shows the fit of the measured conductance to a phenomenological function $\sigma_{xx}=\sigma_0exp[-\Delta E_{eff}/eV_H]$ between $j_x= 1-2~A/m$.
Here $\sigma_0$ is a prefactor which is found to be $0.12\pm 0.03~e^2/h$ and $\Delta E_{eff}=30\pm3~meV$ is the effective energy gap for the activation of the carriers, for the filling factors $\nu=\pm2$ and both of the sample regions measured.
The exponential dependence of conductivity on $\Delta E_{eff}$ for this regime is due to quasi-elastic inter-Landau level scattering (QUILLS) assisted by large local electric field.
This behavior is also observed by Guignard~\textit{et al.} and analyzed within the variable range hopping model via the dependence of the energy gap on the filling factor.\cite{Guignard}
The breakdown can be described within the phenomenological model of quasi-elastic inter-Landau level scattering (QUILLS).
The measured effective gap $\Delta E_{eff}$ is very close to the experimentally observed thermally activated gap value.\cite{Giesbers2}.
The difference between $\hbar\omega_c/2=65~meV$ and $\Delta E_{eff}=30~meV$ can be attributed to potential fluctuations due to charge inhomogeneities.
Spatially resolved measurements has demonstrated the existence of the potential fluctuations with the intrinsic disorder length scale of $\approx 30~nm$ in a monolayer graphene related to electron hole puddles.\cite{Martin}
In a recent work, it has been measured by temperature dependent magneto transport measurements that the potential fluctuations due to the electron hole puddles around the charge neutrality point in graphene is about $20~meV$.\cite{Kurganova}
Although QUILLS model predicts larger breakdown currents than almost all the breakdown experiments due to the relatively large separation between the LL's,
the impurity potential fluctuations in effect reduces the spatial separation between the LL'ss and enhances the tunneling between the localized puddles of compressible and incompressible states.
Within this picture the breakdown behavior of a sample strongly depends on the distribution and the strength of the potential fluctuations which was also proposed by Sing~\textit{et al.}\cite{Singh}

The breakdown behavior becomes more erratic in small samples as the one investigated in this work.
Here we argue that the $\rho_{xx}$ measurements and $R_{xy}$ measurements should be very different in the narrow sample.
In our experiments $\rho_{xx}$ measurements probe the partial areas of the sample
($L_{7-8}\times w =7.5\times 1~\mu m^2$ and $L_{6-7}\times w = 5\times 1~\mu m^2$).
However the Hall measurements probe a much smaller sample area ($w_p \times w=1\times 0.5~\mu m^2$) where $\omega_p$ and $\omega$ are the width of the voltage probe and the width of the sample respectively.
We indeed observe a difference in the breakdown behavior of the longitudinal and Hall resistances.

The deviation of the Hall resistance from its quantized value is known to be related to the longitudinal resistance in 2DEG samples.
This relation has been experimentally analyzed by temperature driven and current driven dissipative regimes.
An empirical relation $\Delta R_{xy} = -s\rho_{xx}$ has been observed in a numerous experiments and various mechanisms has been proposed to explain the behavior.\cite{Yoshihiro,Cage84,Mooij,Furlan} At high temperatures quadratic dependence, $\Delta R_{xy} \propto \rho_{xx}^2$ was also seen.\cite{Furlan}
The observed value of the $s$ in the linear relation varies but typically is in the order of unity.
One of the proposed mechanisms is the geometrical effect, i.e. the mixing of the longitudinal resistivity to Hall resistance when the current density and the Hall field are not orthogonal under dissipative conditions, $\Delta R_{xy} = -(w_p/w)\rho_{xx}$.\cite{Mooij}
In the temperature driven dissipative quantum Hall regime, Cage \textit{et al.} experimentally verified the linear relationship over four orders of magnitude change in resistivity and determined the $s$ values to vary in the range of 0.015-0.5 depending on the sample and the configuration of the probes.\cite{Cage84}
The behavior can be understood by the thermal activation of electrons to higher Landau levels.
The deviation of $R_{xy}$ versus $\rho_{xx}$ was also analyzed when the 2DEG is driven into dissipative regime by increasing the current and again a linear relation is observed. In the current driven dissipation the value of the $s$ and its variation with the filling factor verified the role of the variable range hopping (VRH) mechanism for the activation of the carriers.\cite{Furlan}

We analyzed the current driven deviation of the quantized Hall resistance, $\Delta R_{xy}$ as a function of the longitudinal $\rho_{xx}$ for the filling factor $\nu=2$ as shown in Fig~5.
We observe two regimes in this plot, for the current densities up to 5A/m the deviation follows a linear behavior with a slope, $s = 0.066 \pm 0.008$.
For $j_x = 5-20~A/m$ however the deviation in the Hall resistance shows and abrupt increase and the slope jumps to $s = 0.40$.
The small value of $s$ up to $j_x=5~A/m$ is due to the weak dependence of $R_{xy}$ on the increased current compared to $\rho_{xx}$.
Above $j_x=5 A/m$, $\Delta R_{xy}$ suddenly starts to increase with the current.
This is consistent with the qualitative observation of the different onset of the breakdown in the $\rho_{xx}$ and $R_{xy}$ plots in Fig~3.

\begin{figure}[h]
\begin{center}
 \includegraphics[width=0.45\textwidth]{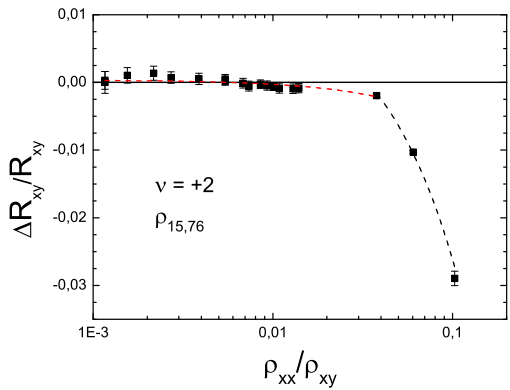}\
 \caption {\label{Fig5}
  Relative deviation of the Hall resistance, $\Delta R_{xy}$/$R_{xy}$ versus the normalized longitudinal resistivity, $\Delta\rho_{xx}^{min}$/$R_{xy}$ at the filling factor $\nu=2$. Semilog scale is used to clearly display all data points. Red and black lines are the linear fits to two range of data corresponding to $j_x \leq 5~A/m$ and $5~A/m~\leq~j_x\leq$~20~A/m respectively.}

\end{center}
\end{figure}

High critical current for the Hall resistance compared to the lower critical breakdown current of longitudinal resistance is caused by the narrow, small and more homogeneous region across the Hall probes 4-6.
In this particular small region of the sample to Hall plateaus remain flat well up to high fields possibly due to slower variation of the potential fluctuations. On the other hand, longitudinal resistance measurements probe wider regions along which more electron-hole puddles exist leading to smaller breakdown currents.
Especially, when the length scale of inhomogeneity becomes comparable with the sample width, inhomogeneity size effect leads to variation in the range and distribution of the localized dissipative regions throughout the sample interior. This leads to variation of the breakdown current throughout the sample area.
Such inhomogeneity effects in the breakdown of the quantum Hall effect were also observed in narrow GaAs 2DEG samples. \cite{Sachrajda}

A similar phenomenon was observed by Kawaji \textit{et al.} in a GaAs 2DEG samples with $35\mu m$ width, "a phase separation between the quantum Hall resistance state and the dissipative quasi-quantum Hall resistance state".\cite{Kawaji} The Hall resistance remained precisely quantised even when the longitudinal resistance was increasing exponentially with the current. Then a \textit{collapse}, a steep change  in the quantized resistance occurred while longitudinal resistance kept its gradual increase. They explained the results using a phenomenological model which assumes that the Hall bar consists of alternating phases of ideal quantum Hall and dissipative quasi-quantum Hall states along the bar. Collapse of quantized Hall resistance occurs when the quantum Hall phase disappears near the voltage probes, hence it is a different phenomenon than breakdown.

\section{Conclusions}

We observed that in a narrow monolayer graphene Hall bar sample the critical current for the breakdown of the quantum Hall effect varies between 0.50 and 0.75~A/m depending on the probe set and the filling factor. The breakdown emerges as a gradual increase in the longitudinal resistivity rather than an abrupt jump and its behavior strongly depends on the location and the gate potential. The Hall resistance also displays a different behavior with the increasing current. Its deviation with current remains very small until an abrupt increase around $j_x= 5~A/m$. The exponential dependence of the longitudinal resistivity on $1/j_x$ is attributed to impurity mediated inter-Landau level tunnelling of carriers. Charge induced inhomogeneity is critical for the breakdown of the quantum Hall effect, especially when the length scale of the fluctuations becomes comparable with the sample size in graphene devices. We interpret that the physical mechanism underlying the occurrence of local breakdown is the highly inhomogeneous distribution of impurity induced potential fluctuations in the sample.

\section{Acknowledgements}

The authors would like to thank Cem Celebi, Anil Gunay-Demirkol, Gorkem Soyumer, Mahmut Tosun, Taylan Erol, Abdulkadir Canatar and Hikmet Coskun for fruitful discussions and their support.
This work was supported by Scientific and Technological Research Council of Turkey (TUBITAK) under Project Grant No. 107T855.

\end{document}